\documentclass[twocolumn,showpacs,preprintnumbers,amsmath,amssymb,nofootinbib,APS]{revtex4}

\usepackage{dcolumn}
\usepackage{bm}
\usepackage[english]{babel}
\usepackage{amsfonts}
\usepackage{amssymb}
\usepackage{graphicx}

\begin{document}

\preprint{gr-qc/0612047}

\title{{\bf Limit to General Relativity in $f(R)$ theories of gravity}}
\author{Gonzalo J. Olmo}\email{olmoalba@uwm.edu}
\affiliation{ {\footnotesize Physics
Department, University of Wisconsin-Milwaukee,Milwaukee, WI 53201,
USA }}

\date{December 7th, 2006}

\begin{abstract}
We discuss two aspects of $f(R)$ theories of gravity in metric formalism. We first study the reasons to introduce a scalar-tensor representation for these theories and the behavior of this representation in the limit to General Relativity, $f(R)\to R$. We find that the scalar-tensor representation is well behaved even in this limit. Then we work out the exact equations for spherically symmetric sources using the original $f(R)$ representation, solve the linearized equations, and compare our results with recent calculations of the literature. We observe that the linearized solutions are strongly affected by the cosmic evolution, which makes very unlikely that the cosmic speedup be due to $f(R)$ models with correcting terms relevant at low curvatures.
\end{abstract}

\pacs{98.80.Es , 04.50.+h, 04.25.Nx}

\maketitle

\section{ Introduction}

In the last few years, modified theories of gravity of the $f(R)$ type, where $R$ is the Ricci scalar, have received much attention. These theories have the ability to generate late-time cosmic acceleration and also early-time inflation depending on the details of the function $f(R)$ considered. In this sense, they represent an
interesting alternative to the combination of General Relativity (GR) plus dark energy models \cite{Pad06,Dark}, and to other explanations for the cosmic speedup \cite{Pad02,Parker}. A given gravity lagrangian $f(R)$ can, in addition, lead to two completely different theories depending on the variational principle used to derive the equations of motion. Here we will be dealing with the most common choice, the metric variational formalism. The other choice, Palatini formalism, assumes that the connection is a field independent of the metric and will not be considered here
(see \cite{Aea06,ABF04,BGS06,CCT05,Koi06,K-K06,L-C06,M-W03,Olmo06,O-K04,Pop06,Sot06b,A-S06} for details). \\

The dominant underlying philosophy in the construction of
$f(R)$ models has been that non-linear curvature terms
that grow at low curvatures could have a negligible effect in
stellar systems or galactic scales, where the curvature is assumed to be relatively high,
but a non-trivial one in the cosmic regime, where the curvatures involved
are orders of magnitude smaller. This idea has led to many different
proposals, among which the Carroll et al. model $f(R)=R-\mu^4/R$
\cite{CDTT03} is perhaps the most known (see also \cite{Capo02} for earlier models).
A rich debate in various arenas has taken place with some authors focusing on the
cosmological aspects\footnote{See for instance \cite{M-W03,N-O03,CDTT05} }, and others on
the predictions of these theories at smaller scales (Newtonian and
post-Newtonian regimes). Among the latter, some claim that models
of the $1/R$ type are ruled out by solar system and/or laboratory
experiments \cite{Chi03,Fla04,Olmo05,Kam06,JLL06,Chi06} (see also \cite{Zak06} for a discussion of $R^n$ models), and others defend the opposite \cite{N-O06,C-T05,AFRT05,R-O06,A-R06,Far06} (see also \cite{Sot06} for a critical discussion of these two positions). An important element in this discussion was the identification of $f(R)$ theories with a class of scalar-tensor theories \cite{T-T83,Chi03,Fla04}, which allowed one to reinterpret the equations of motion and made more accessible the computation of the Newtonian and post-Newtonian limits. More recently, however, this scalar-tensor approach has been criticized and, apparently, shown to break down for theories close to GR \cite{Far06}. It was then claimed that solar system
tests do not actually rule out theories of the type $f(R)=R+\epsilon g(R)$, with $\epsilon$ a small parameter, because the corresponding scalar-tensor theory is not well defined in the limit $\epsilon\to0$.
This conclusion, on the other hand, has also been criticized recently in \cite{Kam06,JLL06,Chi06}, where spherically symmetric weak field configurations have been studied using the original $f(R)$ representation. In those works, it has been found that the post-Newtonian parameter $\gamma$ turns out to be $\gamma=1/2$, in agreement with the results found using the scalar-tensor approach in \cite{Chi03}. Solar system experiments again seem to rule out $f(R)$ theories. However, there is something unclear in those works that should be reanalyzed before accepting their conclusions. In \cite{Kam06,JLL06,Chi06} the {\it wrong} coordinate system was used to carry out the calculations and identify the post-Newtonian parameter $\gamma$. In fact, for the discussion of solar system experiments and the identification of post-Newtonian parameters, one should use isotropic coordinates instead of Schwarzschild coordinates (see \cite{MTW73}, page $1097$ for details). This point and others  will be carefully addressed here.\\

In this work we will discuss in some detail the scalar-tensor representation of $f(R)$ models of the type $f(R)=R+\epsilon g(R)$ and their limit to GR. Firstly, we will motivate the introduction of the scalar-tensor representation and will then show that it is well defined even in the limit $\epsilon \to 0$. Secondly, we will consider the weak field limit for spherically symmetric sources using the original $f(R)$ representation. We will show that the same results found in \cite{Olmo05} using the scalar-tensor approach are found here using the $f(R)$ form. For our calculations we use isotropic coordinates. In the appendix, we also solve the linearized equations using Schwarzschild coordinates and transform the solution to isotropic coordinates. This will serve as a consistency check for our calculations. In order to provide a complete description of the dynamics of $f(R)$ theories, in our derivation we will take into account the interaction of the local system with the background cosmology. This will allow us to better understand the limit to General Relativity of $f(R)$ theories. We will see how, driven by the cosmic expansion, a theory characterized by a low curvature scale $R_\epsilon$ goes from a General Relativistic phase, in which the post-Newtonian parameter $\gamma\approx 1$,  to a scalar-tensor phase, in which $\gamma\to 1/2$. Though we will be mainly discussing the weak field regime, we provide here the basic concepts and formulas that (hopefully) will be used in a forthcoming work \cite{O-U07}, in which the strong field regime and the contribution of higher-order corrections to the linearized theory \cite{N-A06} will be studied.\\

The paper is organized as follows. We first define the action and derive the equations of motion for $f(R)$ theories. Those equations are then physically interpreted, which motivates the introduction of the scalar-tensor representation. We then discuss the conditions for the equivalence between $f(R)$ theories and scalar-tensor theories and study the limit $f(R)\to R$. The next step is to write the equations of motion for spherically symmetric systems and obtain the relevant solutions for weak sources. We then discuss the results obtained and the interaction of the system with the background cosmology. We finish with a brief summary and conclusions. In the appendix we solve the equations using Schwarzschild coordinates as a consistency check for our results.

\section{The theory}\label{sec:theory}

 The action that defines $f(R)$ gravities has the generic form
\begin{equation}\label{eq:def-f(R)}
S=\frac{1}{2\kappa ^2}\int d^4 x\sqrt{-g}f(R)+S_m[g_{\mu
\nu},\psi_m]
\end{equation}
where $S_m[g,\psi_m]$ represents the matter action, which depends
on the metric $g_{\mu \nu }$ and the matter fields $\psi_m$, and $\kappa^2$ is a constant with suitable units\footnote{We will see later how $\kappa^2$ is related to the effective Newton's constant [see Eq.(\ref{eq:G}) below].}. For
notational purposes, we remark that the scalar $R$ is defined as
the contraction $R=g^{\mu \nu }R_{\mu \nu }$, where $R_{\mu \nu }$
is the Ricci tensor
\begin{equation}\label{eq:def-Ricci}
R_{\mu\nu}=-\partial_{\mu}
\Gamma^{\lambda}_{\lambda\nu}+\partial_{\lambda}
\Gamma^{\lambda}_{\mu\nu}+\Gamma^{\lambda}_{\mu\nu}\Gamma^{\rho}_{\rho\lambda}-\Gamma^{\lambda}_{\nu\rho}\Gamma^{\rho}_{\mu\lambda}
\end{equation}
and $\Gamma^\alpha_{\beta \gamma }$ is the connection, which is
defined as
\begin{equation}\label{eq:def-Gamma}
\Gamma^\alpha_{\beta \gamma }=\frac{g^{\alpha \lambda
}}{2}\left(\partial_\beta g_{\lambda \gamma }+\partial_\gamma
g_{\lambda \beta }-\partial_\lambda g_{\beta \gamma }\right)
\end{equation}
Variation of (\ref{eq:def-f(R)}) leads to the following field
equations for the metric
\begin{equation}\label{eq:f-var}
f'(R)R_{\mu\nu}-\frac{1}{2}f(R)g_{\mu\nu}-
\nabla_{\mu}\nabla_{\nu}f'(R)+g_{\mu\nu}\Box f'(R)=\kappa ^2T_{\mu
\nu }
\end{equation}
where $f'(R)\equiv df/dR$. According to (\ref{eq:f-var}), we
see that, in general, the metric satisfies a system of
fourth-order partial differential equations. The trace of
(\ref{eq:f-var}) takes the form
\begin{equation}\label{eq:trace-m}
3\Box f'(R)+Rf'(R)-2f(R)=\kappa ^2T
\end{equation}
This equation will be useful for the physical interpretation of
the field equations.\\

\subsection{Physical interpretation}\label{sec:Physics}

Let us consider a generic $f(R)$ theory not necessarily close to
GR and rewrite (\ref{eq:f-var}) in the form
\begin{eqnarray}\label{eq:Gab-ST}
R_{\mu \nu }-\frac{1}{2}g_{\mu \nu }R&=&
\frac{\kappa^2}{f'(R)}T_{\mu\nu}-\frac{1}{2f'(R)}g_{\mu\nu}[Rf'(R)-f(R)]+\nonumber\\
  &+&\frac{1}{f'(R)}\left[\nabla_\mu\nabla_\nu f'(R)-g_{\mu\nu}\Box f'(R)\right]
\end{eqnarray}
The right hand side of this equation can now be seen as the source
terms for the metric. This equation, therefore, tells us that the
metric is generated by the matter and by terms related to the
scalar curvature. If we now wonder about what generates the scalar
curvature, the answer is in (\ref{eq:trace-m}). That expression
says that the scalar curvature satisfies a second-order
differential equation with the trace $T$ of the energy-momentum
tensor of the matter and other curvature terms acting as sources.
We have thus clarified the role of the higher-order derivative
terms present in (\ref{eq:f-var}). The scalar curvature is now
a dynamical entity which helps generate the space-time metric and
whose dynamics is determined by (\ref{eq:trace-m}).\\

At this point one should have noted the essential
difference between a generic $f(R)$ theory and GR. In GR the only
dynamical field is the metric and its form is fully characterized
by the matter distribution through the equations
$G_{\mu\nu}=\kappa^2T_{\mu\nu}$. The scalar curvature is also
determined by the local matter distribution but through an
algebraic equation, namely, $R=-\kappa^2T$. In the $f(R)$ case
both $g_{\mu\nu}$ and $R$ are dynamical fields, i.e., they are
governed by differential equations. Furthermore,  the scalar
curvature $R$, which can obviously be expressed in terms of the
metric and its derivatives using (\ref{eq:def-Ricci}), now
plays a non-trivial role in the determination of the metric itself.\\

\subsection{ Scalar-Tensor representation} \noindent

The physical interpretation given above puts forward the central
and active role played by the scalar curvature in the field
equations of $f(R)$ theories. However, (\ref{eq:trace-m})
suggests that the actual dynamical entity is $f'(R)$ rather than
$R$ itself. This is so because, besides the metric, $f'(R)$ is the
only object acted on by differential operators in the field
equations. Motivated by this, we can introduce the following
alternative notation
\begin{eqnarray}\label{eq:phi=f'}
\phi&\equiv& f'(R)\\
V(\phi)&\equiv& R(\phi)f'-f(R(\phi)) \label{eq:V=rf'-f}
\end{eqnarray}
and rewrite (\ref{eq:f-var}) as
\begin{eqnarray}\label{eq:Gab-phi}
R_{\mu \nu }(g)-\frac{1}{2}g_{\mu \nu }R(g)&=&
\frac{\kappa^2}{\phi}T_{\mu\nu}-\frac{1}{2\phi}g_{\mu\nu}V(\phi)+\nonumber\\
  &+&\frac{1}{\phi}\left[\nabla_\mu\nabla_\nu\phi-g_{\mu\nu}\Box \phi\right]
\end{eqnarray}
Using the same notation, (\ref{eq:trace-m}) turns into
\begin{equation}\label{eq:phi-ST}
3\Box \phi +2V(\phi)-\phi \frac{dV}{d\phi}=\kappa^2T
\end{equation}
This slight change of notation helps us identify
(\ref{eq:Gab-phi}) and (\ref{eq:phi-ST}) with the field
equations of a Brans-Dicke theory with parameter $\omega=0$ and
non-trivial potential $V(\phi)$, whose action takes the form
\begin{equation}\label{eq:BD0}
S=\frac{1}{2\kappa ^2}\int d^4 x\sqrt{-g}[\phi R
-V(\phi)]+S_m[g_{\mu \nu},\psi_m]
\end{equation}
In terms of this scalar-tensor representation our interpretation
of the field equations of $f(R)$ theories is obvious, since both
the matter and the scalar field help generate the metric. The
scalar field is a dynamical object influenced by the matter and by
self-interactions according to (\ref{eq:phi-ST}).\\

We would like to remark that, according to the above derivation,
the only requirement needed to express the field equations of an
$f(R)$ theory in the form of a Brans-Dicke theory is that $f'(R)$
be invertible\footnote{The minimum requirement for a function of
one variable to be invertible in an interval is that the function
be continuous and one-to-one in that interval.}, i.e., that $R(f')$ exist. This
is necessary for the construction of $V(\phi)$. Therefore, the
differentiability condition $f''(R)\neq 0$ that one finds
following other derivations of the scalar-tensor representation
\cite{T-T83,Chi03,Fla04,Far06} is just a superfluous condition generated by the particular
method used\footnote{If $f''(R)$ vanishes at some point $R_0$, we may need to choose
among different branches to extend the solutions $R(f')$ beyond $R_0$. In other words, $R(f')$ may not be unique in those cases.}.\\
The inverse problem of finding the $f(R)$ theory corresponding to
a given scalar-tensor theory of the form given in
(\ref{eq:BD0}) also requires an invertibility condition only. In
this case, the equations of motion lead to $R=dV/d\phi$. If this
expression can be inverted to obtain $\phi(R)$ then the
corresponding $f(R)$ lagrangian is given by
\begin{equation}
f(R)=R\phi(R)-V(\phi(R))
\end{equation}
We thus conclude that the condition for the equations of motion of
an $f(R)$ theory to be equivalent to those of a $\omega=0$
Brans-Dicke theory is that the function $f'(R)$ be invertible.
Conversely, for a given $\omega=0$ Brans-Dicke theory the
condition for the equivalent $f(R)$ theory to exist is that the
function $dV(\phi)/d\phi=R$ be invertible.

\subsection{Limit $f(R)\to R$}

Now that the scalar-tensor representation of $f(R)$ theories has
been introduced and the conditions for their equivalence
clarified, we will consider the limit $f(R)$ going to $R$ to gain
some insight on their dynamical properties. We will
also show that the equivalence is, by no means, broken in this
limit.\\
Let us concentrate on models of the form $f(R)=R+\epsilon g(R)$
with $\epsilon$ an adjustable small parameter. The scalar field
is therefore identified with $\phi=1+\epsilon g'(R)$ and becomes a
constant, $\phi=1$, in the limit $\epsilon\to 0$. The relevant
part of $\phi$ is therefore contained in $\varphi=g'(R)$. By
inverting this relation we find $R=R(\varphi)$, which allows us to
express the scalar potential as
$V(\phi)=\epsilon[R(\varphi)\varphi-g(R[\varphi])]\equiv
\epsilon\tilde{V}(\varphi)$. The action (\ref{eq:BD0}) then
turns into
\begin{equation}
S=\frac{1}{2\kappa^2}\int d^4 x\sqrt{-g}\left[(1+\epsilon\varphi)
R-\epsilon\tilde{V}(\varphi)\right]+S_m[g_{\mu \nu},\psi_m]
\end{equation}
As expected, the limit $\epsilon\to 0$ leads smoothly to the
action of GR. Let us now look at the equation of motion for $\phi$
to see what happens in that limit. In terms of $\varphi$,
(\ref{eq:phi-ST}) can be recast as
\begin{equation}\label{eq:phi-epsilon}
\epsilon\left[3\Box \varphi +\varphi R(\varphi)-2
g(R)\right]=R(\varphi)+\kappa^2T
\end{equation}
In the limit $\epsilon\to 0$ the dynamical part of this equation, that involving
$\Box\varphi$, vanishes and we recover the familiar (algebraic) relation
$R=-\kappa^2T$. This means that $\varphi$ has completely decoupled
from the theory. As a consequence $R$ becomes independent of
$\varphi$ and we can state with confidence that the limit to GR is
smooth. This shows that the equivalence between $f(R)$ gravities
and their related Brans-Dicke theories holds even in the limit of GR, since no
inconsistency is found in this limit.\\

To conclude this section, we mention that from
(\ref{eq:phi-epsilon}) we can extract valuable physical
information. The parameter $\epsilon$, which we now assume to be small and fixed, in front of the $\Box
\varphi$ term makes apparent the existence of two regimes. The
first one corresponds to a behavior very close to GR, in which
$R\approx-\kappa^2T$ is a good approximation. The second regime
arises when the terms on the right hand side are small. In that
situation, the contribution from the left hand side can no
longer be neglected. The behaviour of $R$ is then dominated by
$\varphi$, i.e., the full dynamics of (\ref{eq:phi-epsilon})
must be taken into account, which means that $R$ receives contributions
from $T$ and from its self-interactions (or, equivalently, from
$\varphi$). In this situation, the theory behaves as a scalar-tensor theory, with
$\varphi$ representing the scalar degree of freedom. We will discuss further this point later on.\\

\section{Applications}\label{sec:calculations}

A complete description of a physical system must take into account not only the system but also its interaction with the environment. In this sense, any physical system is surrounded by the rest of the universe. The relation of the local system with the rest of the universe manifests itself in a set of boundary conditions. In our case, according to (\ref{eq:trace-m}) and (\ref{eq:Gab-ST}), the metric and the function $f'$ (or, equivalently, $R$ or $\phi$) are subject to boundary conditions, since they are dynamical fields (they are governed by differential equations). The boundary conditions for the metric can be trivialized by a suitable choice of coordinates. In other words, we can make the metric Minkowskian in the asymptotic region and fix its first derivatives to zero (see chapter 4 of \cite{Wil93} for details). The function $f'(R)$, on the other hand, should tend to the cosmic value $f'(R_c)$ as we move away from the local system. The precise value of $f'(R_c)$ is obtained by solving the equations of motion for the corresponding cosmology. According to this, the local system will interact with the asymptotic (or background) cosmology via the boundary value $f'(R_c)$ and its cosmic-time derivative. Since the cosmic time-scale is  much larger than the typical time-scale of local systems (billions of years versus years), we can assume an adiabatic interaction between the local system and the background cosmology. We can thus neglect terms such as $\dot{f}'(R_c)$, where dot denotes derivative with respect to the cosmic time.\\
The problem of finding solutions for the local system, therefore, reduces to solving (\ref{eq:Gab-ST}) expanding about the Minkowski metric in the asymptotic region\footnote{Note that the expansion about the Minkowski metric does not imply the existence of global Minkowskian solutions. As we will see, the general solutions to our problem turn out to be asymptotically de Sitter spacetimes.}, and (\ref{eq:trace-m}) tending  to
\begin{equation}\label{eq:trace-cosmo}
3\Box_c f'(R_c)+R_cf'(R_c)-2f(R_c)=\kappa ^2T_c
\end{equation}
where the subscript $c$ denotes cosmic value, far away from the system. In particular, if we consider a weakly gravitating local system, we can take $f'=f'_c+\varphi(x)$ and $g_{\mu\nu}=\eta_{\mu\nu}+h_{\mu\nu}$, with $|\varphi|\ll|f'_c|$ and $|h_{\mu\nu}|\ll 1$ satisfying  $\varphi\to 0$ and $h_{\mu\nu}\to 0$ in the asymptotic region. Note that should the local system represent a strongly gravitating system such as a neutron star or a black hole, the perturbative expansion would not be sufficient everywhere. In such cases, the perturbative approach would only be valid in the far region. Nonetheless, the decomposition $f'=f'_c+\varphi(x)$ is still very useful because the equation for the local deviation $\varphi(x)$ can be written as
\begin{equation}\label{eq:trace-local}
3\Box \varphi +W(f'_c+\varphi)-W(f'_c)=\kappa^2T,
\end{equation}
where $T$ represents the trace of the local sources, we have defined $W(f')\equiv R(f')f'-2f(R[f'])$, and $W(f'_c) $ is a slowly changing constant within the adiabatic approximation. In this case, $\varphi$ needs not be small compared to $f'_c$ everywhere, only in the asymptotic regions.

\subsection{Spherically symmetric solutions}

The analysis of spherically symmetric solutions can be carried out in several ways. One of them is to write the line element using Schwarzschild coordinates
\begin{equation}\label{eq:ds2-sch}
ds^2=-B(\tilde{r})dt^2+\frac{1}{C(\tilde{r})}d\tilde{r}^2+\tilde{r}^2d\Omega^2
\end{equation}
Another possibility, more suitable for the discussion of observable effects, is to use isotropic coordinates \cite{MTW73}. We will follow here this second option and, for completeness, we will discuss the other in the appendix to check the consistency of our calculations. We define the line element
\begin{equation}\label{eq:ds2-iso}
ds^2=-A({r})e^{2\psi({r})}dt^2+\frac{1}{A({r})}\left(d{r}^2+{r}^2d\Omega^2\right),
\end{equation}
which, assuming a perfect fluid for the sources, leads to the following field equations (see \cite{M-V06} for a different approach)
\begin{eqnarray}\label{eq:Ar-iso}
A_{rr}+A_r\left[\frac{2}{r}-\frac{5}{4}\frac{A_r}{A}\right]&=&\frac{\kappa^2\rho}{f'}+\frac{R
f'-f(R)}{2f'}+\\ &+& \frac{A}{f'}\left[f'_{rr}+f'_r\left(\frac{2}{r}-\frac{A_r}{2A}\right)\right]\nonumber\\
A\psi_r\left[\frac{2}{r}+\frac{f'_r}{f'}-\frac{A_r}{A}\right]-\frac{A^2_r}{4A}&=&\frac{\kappa^2P}{f'}-\frac{Rf'-f(R)}{2f'}-\nonumber
\\&-& A\frac{f'_r}{f'}\left[\frac{2}{r}-\frac{A_r}{2A}\right]\label{eq:psir-iso}
\end{eqnarray}
where $f'=f'_c+\varphi$, and the subscripts in
$\psi_r, f'_r,f'_{rr},M_r$ denote derivation with respect to the
radial coordinate. Note also that $f'_r=\varphi_r$ and $f'_{rr}=\varphi_{rr}$.
The equation for $\varphi$ is, according to (\ref{eq:trace-local}) and (\ref{eq:ds2-iso}),
\begin{eqnarray} \label{eq:frr-iso}
A\varphi_{rr}&=&-A\left(\frac{2}{r}+\psi_r\right)\varphi_r -\\ &-&
\frac{W(f'_c+\varphi)-W(f'_c)}{3}+\frac{\kappa^2}{3}(3P-\rho)\nonumber
\end{eqnarray}
Equations (\ref{eq:Ar-iso}), (\ref{eq:psir-iso}),  and (\ref{eq:frr-iso}) can be used to work out the metric of any spherically symmetric system regardless of the intensity of the gravitational field. For weak sources, however, it is convenient to expand them assuming $|\varphi|\ll f'_c$ and $A=1-2M(r)/r$, with $2M(r)/r\ll1$. The result is
\begin{eqnarray}\label{eq:Ar-iso-lin}
-\frac{2}{r}M_{rr}(r)&=&\frac{\kappa^2\rho}{f'_c}+V_c+\frac{1}{f'_c}\left[\varphi_{rr}+\frac{2}{r}\varphi_r\right]\\
\frac{2}{r}\left[\psi_r+\frac{\varphi_r}{f'_c}\right]&=&\frac{\kappa^2}{f'_c}P-V_c\label{eq:psir-iso-lin}\\
\varphi_{rr}+\frac{2}{r}\varphi_r&=&\frac{
\kappa^2}{3}(3P-\rho)+m_c^2\varphi
\label{eq:frr-iso-lin}
\end{eqnarray}
where we have defined
\begin{eqnarray}\label{eq:Vc}
V_c&\equiv& \frac{R_cf'_c-f(R_c)}{2f'_c}\\ \label{eq:mass}
m^2_c&=&\frac{f'(R_c)-R_cf''(R_c)}{3f''(R_c)}
\end{eqnarray}
This expression for $m^2_c$ was first found in \cite{Olmo05} within the scalar-tensor approach. It was found there that $m^2_c>0$ is needed to have a well-behaved (non-oscillating) Newtonian limit. This expression and the conclusion $m^2_c>0$ were also reached in \cite{F-N05} by studying the stability of de Sitter space. The same expression has also been found more recently in \cite{Chi06,N-A06}.\\
Outside of the sources, the solutions of (\ref{eq:Ar-iso-lin}), (\ref{eq:psir-iso-lin}) and (\ref{eq:frr-iso-lin}) lead to
\begin{eqnarray}
\varphi(r)&=&\frac{C_1}{r}e^{-m_c r}\label{eq:phir-iso-sol}\\
A(r)&=&1-\frac{C_2}{r}\left(1-\frac{C_1}{C_2 f'_c}e^{-m_c r}\right)+\frac{V_c}{6}r^2\label{eq:Ar-iso-sol}\\
A(r)e^{2\psi}&=& 1-\frac{C_2}{r}\left(1+\frac{C_1}{C_2 f'_c}e^{-m_c r}\right)-\frac{V_c}{3}r^2\label{eq:psir-iso-sol}
\end{eqnarray}
where an integration constant $\psi_0$ has been absorbed in a redefinition of the time coordinate. The above solutions coincide, as expected, with those found in \cite{Olmo05} for the Newtonian and post-Newtonian limits using the scalar-tensor representation. Comparing our solutions with those, we identify
\begin{eqnarray}
C_2&\equiv& \frac{\kappa^2}{4\pi f'_c}M_\odot\\
\frac{C_1}{f'_cC_2}&\equiv&\frac{1}{3}\label{eq:C1}
\end{eqnarray}
where $M_\odot=\int d^3x \rho(x)$. The line element (\ref{eq:ds2-iso}) can be written as
\begin{eqnarray}\label{eq:ds2-iso-sol}
ds^2&=&-\left(1-\frac{2G M_\odot}{r}-\frac{V_c}{3}r^2\right)dt^2+\\ &+& \left(1+\frac{2G\gamma M_\odot}{r}-\frac{V_c}{6}r^2\right)(dr^2+r^2d\Omega^2)\nonumber
\end{eqnarray}
where we have defined the effective Newton's constant
\begin{equation}\label{eq:G}
G=\frac{\kappa^2}{8\pi f'_c}\left(1+\frac{e^{-m_cr}}{3}\right)
\end{equation}
and the effective post-Newtonian parameter
\begin{equation}\label{eq:gamma}
\gamma=\frac{3-e^{-m_cr}}{3+e^{-m_cr}}
\end{equation}
This completes the lowest-order solution in isotropic coordinates. The higher order corrections to this solution
will be studied in detail elsewhere \cite{O-U07} (see also \cite{N-A06} for a recent discussion of those corrections).\\

Had we used the Schwarzschild line element (\ref{eq:ds2-sch}), the result would be (see the appendix for details)
\begin{eqnarray}\label{eq:ds2-sch-sol}
ds^2&=&-\left(1-\frac{C_2}{\tilde{r}}\left[1+\frac{e^{-m_c\tilde{r}}}{3}\right]-\frac{V_c}{3}\tilde{r}^2\right)dt^2+\\
&+&\left(1+\frac{C_2}{\tilde{r}}\left[1-\frac{1+m_c\tilde{r}}{3}e^{-m_c\tilde{r}}\right]+\frac{V_c}{3}\tilde{r}^2\right)d\tilde{r}^2+\tilde{r}^2d\Omega^2 \nonumber
\end{eqnarray}
We see that the Newtonian limit, the function in front of $dt^2$, coincides with the result obtained using isotropic coordinates (neglecting higher order corrections due to the change of coordinates). However, the first post-Newtonian correction is not correctly identified if we just take the function in front of $d\tilde{r}^2$. It is necessary to transform to isotropic coordinates using
\begin{equation}\label{eq:sch2iso}
\tilde{r}\approx r+\frac{C_2}{2}\left[1-\frac{e^{-m_c r}}{3}\right]-\frac{V_c}{12}r^3
\end{equation}
to find the correct expression. Nonetheless, in the limit $m_cr\ll1$ considered in \cite{Kam06,JLL06,Chi06} and neglecting the $V_cr^2$ terms, the change of coordinates is trivial ($d\tilde{r}/dr\approx 1$) and (\ref{eq:ds2-sch-sol}) agrees with (\ref{eq:ds2-iso-sol}). This is, however, just an accident that leads to the desired result. In general, one should use isotropic coordinates for the correct identification of the post-Newtonian parameters and take into account the interaction with the background cosmology for the right interpretation of the limit to General Relativity (see next subsection).

\subsection{Discussion of the linearized solutions}

We see from (\ref{eq:G}) and (\ref{eq:gamma}) that the parameters $G$ and $\gamma$ that characterize the linearized metric depend on the effective mass $m_c$ (or inverse length scale $m_c^{-1}$). Newton's constant, in addition, also depends on $f'_c$. Since the value of the background cosmic curvature $R_c$ changes with the cosmic expansion, it follows that $f'_c$ and $m_c$ will also change. The variation in time of $f'_c$ will induce a time variation in the effective Newton's constant analogous to the well-known time dependence that arises in Brans-Dicke theories. Actually, if we use the scalar-tensor representation, the analogy becomes an identity. The length scale $m_c^{-1}$, characteristic of $f(R)$ theories, does not appears in Brans-Dicke theories because in the latter the scalar potential is  $V(\phi)\equiv 0$, in contrast with (\ref{eq:V=rf'-f}). \\

Let us now review how the cosmic expansion proceeds in $f(R)$ models with curvature terms that grow at low curvatures (see also \cite{CDTT03,Dick04,Olmo05} for more details). During the matter dominated era, the cosmic expansion takes place as in GR. This is so because $R_c$ and $T_c$ are well above the characteristic scale $R_\epsilon$ of the non-linear terms in the lagrangian. In this situation, we find $R_c\approx -\kappa^2T_c$, as discussed in the paragraph following equation (\ref{eq:phi-epsilon}). At later times, the sources become very diluted and $R_c$ approaches the scale $R_\epsilon$ in which the left hand side of (\ref{eq:phi-epsilon}) can no longer be neglected. The scalar curvature then becomes a fully dynamical object and the theory gets into the scalar-tensor dynamical regime, which triggers the cosmic speedup. Therefore, the cosmic expansion drives the theory from a General Relativistic (decelerated) phase to a scalar-tensor (accelerated) phase. Let us now consider what happens in the solar system. During the General Relativistic phase, characterized by $R_c\gg R_\epsilon, \ f'_c\to 1, \ f''_c\to 0$, we find that $m_c\gg 1$, which leads to $e^{-m_cr}\to 0$ rapidly (very short scalar interaction range). During this period of time, we have $\gamma\to 1$, which coincides with the General Relativistic result and agrees with current observations. At later cosmic times, when $R_c\sim R_\epsilon, \ f''_c>0$, the effective mass $m_c$ becomes finite, decays with time as $f''_c$ grows, and eventually leads to $e^{-m_cr}\approx 1$ over solar system scales (very long scalar interaction range). We then find $\gamma\to 1/2$, which coincides with the value corresponding to the $w=0$ Brans-Dicke theory and is ruled out by observations.\\

The change of dynamical regime in local systems can also be seen by looking at the value of $R$. To illustrate this point, we will use the solutions found in \cite{Olmo05}, which are also valid within the sources. To lowest order, we find
\begin{equation}\label{eq:R}
R=V_c+\frac{m_c^2\kappa^2}{4\pi f'_c}\int d^3x'\frac{\rho(t,\ \vec{x}')}{|\vec{x}-\vec{x}'|}e^{-m_c|\vec{x}-\vec{x}'|}
\end{equation}
It is not difficult to show (considering a small volume centered about $|\vec{x}-\vec{x}'|\to 0$ and integrating by parts) that this expression becomes $R=\kappa^2\rho(t,\vec{x})$ in the limit of General Relativity ($R_c\gg R_\epsilon$, $f'_c\to1, \ f''_c\to0, \ m^2_c\to\infty$, and $\ V_c\to 0$). As the cosmic expansion takes place and $R_c$ approaches the scale $R_\epsilon$ of the scalar-tensor dynamical regime  ($f'_c\neq 1$ , $f''_c>0$ and growing, $m_c^2\to$ finite and decreasing) the dependence of $R$ on the local matter density is softened by the integration in (\ref{eq:R}). At a given point $\vec{x}$, $R$ is then an average over nearby sources with weight $e^{-m_c|\vec{x}-\vec{x}'|}/|\vec{x}-\vec{x}'|$ . This shows that the dependence of $R$ on the local matter density is less and less important as $m_c^2$ decreases, or equivalently, as the interaction range $l_c=m_c^{-1}$ of the scalar field grows due to the cosmic expansion. Note that though the magnitude of $R$ is reduced at the location of the sources, with respect to the value $\kappa^2\rho$ in the limit of GR, it becomes non-zero at farther distances from them. In fact, away from the sources, (\ref{eq:R}) can be approximated by
\begin{equation}\label{eq:R-app}
R=V_c+\frac{m_c^2\kappa^2}{4\pi f'_c}\frac{M_\odot e^{-m_c r}}{r}
\end{equation}
which has a smooth $1/r$ decay when $m_cr\ll 1$. Note that in the limit of infinite range, $m_c\to0$, the curvature $R$ does not depend on the local sources. This shows explicitly that the phylosophy that motivated the design of $f(R)$ models (see the introduction) was wrong, since $R$ not always behaves as in GR. \\

In summary, we have seen that any theory characterized by non-linear terms that grow at low curvatures will evolve from GR during the matter dominated era to a scalar-tensor theory at later times. This transformation takes place at both large (cosmic) and short (solar sytem) scales.  Therefore, if the cosmic speedup were due to the growth of  non-linear corrections in the lagrangian, today we should be in the scalar-tensor phase, which is characterized by $\gamma\to 1/2$. This fact is in conflict with solar system observations ($\gamma_{exp}=1+(2.1\pm 2.3)\cdot 10^{-5}$, \cite{BIT03}) and makes it very unlikely that the cosmic speedup be due to such corrections in the gravity lagrangian.\\

\section{Summary and conclusions}

In this work we have studied two aspects of modified $f(R)$ theories of gravity, namely, their equivalence with $w=0$ Brans-Dicke theories and the weak field solutions in systems with spherical symmetry using the $f(R)$ representation.  We found that the equivalence between $f(R)$ theories and $\omega=0$ Brans-Dicke theories only requires invertibility of the functions $f'(R)$ and $dV/d\phi$. This result puts forward that other well-known derivations of the scalar-tensor representation \cite{T-T83,Chi03,Fla04} introduce an artificial condition, $f''(R)\neq 0$, that may lead to wrong conclusions \cite{Far06} in the limit $\epsilon\to 0$ in models of the form $f(R)=R+\epsilon g(R)$. We explicitly showed that such limit is smooth and that one recovers GR from the scalar-tensor theory.\\
We then worked out the spherically symmetric solutions of the linearized field equations. We carried out the analysis idependently using two different sets of coordinates, namely isotropic and also Schwarzschild coordinates. The results are in perfect agreement when one transforms from one coordinate system to the other, which confirms the validity of our results. We have also made emphasis on the adiabatic interaction between local systems and the background cosmology via boundary conditions. We have seen that the parameters that characterize the linearized solutions depend on the form of the lagrangian and on the value of the cosmic curvature $R_c$ at a given cosmic time. It is then straightforward to see that the GR solution is recovered in the limit $R_c\gg R_\epsilon$, where $R_\epsilon$ represents the characteristic low curvature scale of the lagrangian. When $R_c$, driven by the cosmic expansion, approaches the scale $R_\epsilon$, there is a change of dynamical regime and the theory becomes closer to a scalar-tensor theory than to GR. This change makes the predictions of the theory incompatible with solar system observations, since $\gamma$ goes from $\gamma\sim 1$ in the GR phase to $\gamma\to1/2$ in the late stages of the scalar-tensor phase. Therefore, since the same mechanism that predicts cosmic speedup is also responsible for the running of the parameter $\gamma$, it seems very unlikely that this type of $f(R)$ theories may be responsible for the observed cosmic acceleration.

\appendix

\section{Schwarzschild coordinates}

Using a line element of the Schwarzschild form
\begin{equation}\label{eq:ds2}
ds^2=-B(\tilde{r})dt^2+\frac{1}{D(\tilde{r})}d\tilde{r}^2+\tilde{r}^2d\Omega^2
\end{equation}
the equations that follow from (\ref{eq:Gab-ST}), assuming a perfect fluid, are
\begin{eqnarray}\label{eq:Dr}
\left[\frac{2}{\tilde{r}}+\frac{f'_{\tilde{r}}}{f'}\right]\frac{D_{\tilde{r}}}{2}-\frac{1-D}{r^2} &=&-
\left[\frac{\kappa^2}{f'}\rho+\frac{R
f'-f(R)}{2f'}+\right.\nonumber\\ &+& \left.\frac{D}{f'}\left(f'_{\tilde{r}\tilde{r}}+\frac{2}{r}f'_{\tilde{r}}\right)\right]\\
\frac{D}{2}\left[\frac{2}{\tilde{r}}+\frac{f'_{\tilde{r}}}{f'}\right]\frac{B_{\tilde{r}}}{B}-\frac{1-D}{r^2}&=&
\frac{\kappa^2}{f'}P-\frac{R
f'-f(R)}{2f'}-\nonumber\\ &-& \frac{2D}{\tilde{r}}\frac{f'_{\tilde{r}}}{f'}\label{eq:Br}
\end{eqnarray}
where $f'=f'_c+\varphi$, and the subscripts in
$D_{\tilde{r}} , B_{\tilde{r}}, f'_{\tilde{r}} , f'_{\tilde{r}\tilde{r}}$ denote derivation with respect to the
radial coordinate. Note also that $f'_{\tilde{r}}=\varphi_{\tilde{r}}$ and $f'_{\tilde{r}\tilde{r}}=\varphi_{\tilde{r}\tilde{r}}$.
The equation for $\varphi$ is, according to (\ref{eq:trace-local}) and (\ref{eq:ds2}),
\begin{eqnarray} \label{eq:frr}
D\varphi_{\tilde{r}\tilde{r}}&=& -D\left[\frac{2}{\tilde{r}}+\frac{B_{\tilde{r}}}{2B}-\frac{D_{\tilde{r}}}{2D}\right]\varphi_{\tilde{r}}-\\
&-&
\frac{W(f'_c+\varphi)-W(f'_c)}{3}+\frac{\kappa^2}{3}\left(3P-\rho\right)\nonumber
\end{eqnarray}
For weak sources it is convenient to expand (\ref{eq:Dr}), (\ref{eq:Br}),  and (\ref{eq:frr})  assuming $|\varphi|\ll f'_c$, $D(\tilde{r})=1-2M(\tilde{r})/\tilde{r}$, with $2M(\tilde{r})/\tilde{r}\ll 1$, and $B(\tilde{r})=1-\Phi(\tilde{r})$, with $\Phi(\tilde{r})\ll1$. The result is
\begin{eqnarray}\label{eq:Dr-lin}
\frac{2}{\tilde{r}^2}M_{\tilde{r}} &=&
\frac{\kappa^2}{f'_c}\rho+V_c+\frac{1}{f'_c}\left(\varphi_{\tilde{r}\tilde{r}}+\frac{2}{\tilde{r}}\varphi_{\tilde{r}}\right)\\
\Phi_{\tilde{r}}&=&-\frac{2M(\tilde{r})}{\tilde{r}^2}+V_c\tilde{r}+\frac{2\varphi_{\tilde{r}}}{f'_c}\label{eq:Br-lin}\\
\varphi_{\tilde{r}\tilde{r}}&=& -\frac{2}{\tilde{r}}\varphi_{\tilde{r}}+ m^2_c\varphi-\frac{\kappa^2}{3f'_c}\rho \label{eq:frr-lin}
\end{eqnarray}
where $V_c$ and $m^2_c$ have the same definitions as in the isotropic case. The solutions to the above equations lead to
\begin{eqnarray}
\varphi(\tilde{r})&=& \frac{\tilde{C}_1e^{-m_c\tilde{r}}}{\tilde{r}}\label{eq:frr-assol}\\
M(r)&=&
\tilde{C}_2-\frac{\tilde{C}_1}{2f'_c}[1+m_c\tilde{r}]e^{-m_c\tilde{r}}+\frac{V_c}{6}\tilde{r}^3 \label{eq:Mr-assol}\\
\Phi(\tilde{r})&=&\frac{2\tilde{C}_2}{\tilde{r}}\left[1+\frac{\tilde{C}_1}{2\tilde{C}_2f'_c}e^{-m_c\tilde{r}}\right]+\frac{V_c}{3}\tilde{r}^2
\label{eq:psir-assol}
\end{eqnarray}
Comparing the Newtonian potential $\Phi(\tilde{r})$ with the isotropic solution, we identify $C_2=2\tilde{C}_2, \ C_1=\tilde{C}_1$. A transformation from Schwarzschild to isotropic coordinates [see (\ref{eq:sch2iso})] shows that this identification is consistent with the first post-Newtonian correction, and confirms the validity of our calculations.

\acknowledgments
This work has been supported by NSF grants PHY-0071044 and PHY-0503366. The author thanks Koji Uryu for very useful discussions.

\end{document}